\documentclass[12pt,onecolumn,journal,draftcls,twoside]{IEEEtranTCOM}
\usepackage[english]{babel}
\usepackage{xcolor}
\usepackage[all]{xy}
\usepackage{graphicx}
\usepackage{subfigure}
\usepackage{url}
\usepackage{subfigure}
\usepackage{amsthm}
\usepackage{amsmath}
\usepackage{amssymb}
\usepackage{multirow}
\usepackage{algorithm}
\usepackage{mathrsfs}
\usepackage{url}

\DeclareGraphicsExtensions{.pdf}

\DeclareMathOperator*{\argmax}{arg\,max}
\newcommand{\supp}{\mbox{supp}}
\newcommand{\sgn}{\mbox{sgn}}
\newcommand {\abs}[1]{\left\vert{#1}\right\vert}

\newcommand{\mrm}{\mathrm}

\newtheorem{definition}{Definition}

\newtheorem{conjecture}{Conjecture}

\newcommand \void{\hbox{\rm \small \O}}

\begin{document}
\title{Finite Alphabet Iterative Decoders, Part II: Improved Guaranteed Error
Correction of LDPC Codes via Iterative Decoder Diversity}

\author{David Declercq,~\IEEEmembership{Senior~Member,~IEEE},
Bane~Vasi$\acute{\mathrm{c}}$,~\IEEEmembership{Fellow,~IEEE},
\\ Shiva Kumar Planjery,~\IEEEmembership{Student~Member,~IEEE}, and Erbao Li
\thanks{This work was funded by the NSF under grant CCF-0963726 and the Institut
Universitaire de France grant. Part of the material in this paper was presented
at the Information theory and Applications workshop (ITA'2012).}

\thanks{D.~Declercq and E.~Li are with ENSEA/University of Cergy-Pontoise/CNRS
UMR 8051, 95014 Cergy-Pontoise, France (email: \{declercq,erbao.li\}@ensea.fr)}
\thanks{B.~Vasi$\acute{\mathrm{c}}$ is with the Department of Electrical and
Computer Engineering, University of Arizona, Tucson, AZ, 85719 USA (e-mail:
vasic@ece.arizona.edu).}
\thanks{S.~K.~Planjery is with both the above institutions (email:
shivap@ece.arizona.edu).}
}
\maketitle

\begin{abstract}
Recently, we introduced a new class of finite alphabet iterative decoders
(FAIDs) for low-density parity-check (LDPC) codes. These decoders are capable of
surpassing belief propagation in the error floor region on the Binary Symmetric
channel with much lower complexity. In this paper, we introduce a a novel scheme
to further increase the guaranteed error correction capability from what is
achievable by a FAID on column-weight-three LDPC codes. The proposed scheme uses
a plurality of FAIDs which collectively correct more error patterns than a
single FAID on a given code. The collection of FAIDs utilized by the scheme is
judiciously chosen to ensure that individual decoders have different decoding
dynamics and correct different error patterns. Consequently, they can
collectively correct a diverse set of error patterns, which is referred to as
decoder diversity. We provide a systematic method to generate the set of FAIDs
for decoder diversity on a given code based on the knowledge of the most harmful
trapping sets present in the code. Using the well-known column-weight-three
$(155,64)$ Tanner code with $d_{min}$ = 20 as an example, we describe the method
in detail and show that the guaranteed error correction capability can be
significantly increased with decoder diversity. 
\end{abstract}

\section{Introduction}
It is now well established that iterative decoding based on belief propagation
(BP) approaches the performance of maximum likelihood decoding (MLD) of the low
density parity check (LDPC) codes asymptotically in the block length. However,
for finite length LDPC codes, the sub-optimality of iterative decoding manifests
itself as the inability of the decoder to correct some low-noise configurations
due to the presence of specific subgraphs in the Tanner graphs of the code,
generically termed as trapping sets \cite{richardsonerrorfloors,ontology}. The
presence of trapping sets in a code gives rise to the error floor phenomenon
which is an abrupt degradation in the error rate performance of the code in the
high signal to noise ratio regime. This performance degradation has also been
characterized by the notion of pseudo-codewords \cite{pseudocodewords}, which
represent attractor points of iterative message passing decoders, analogous to
codewords which are the solutions of the MLD. A precise structural relationship
between trapping sets and pseudo-codewords of a given Tanner graph and a
decoding algorithm is not yet fully established, but it has been observed that
the supports of pseudo-codewords are typically contained in small topological
structures of the LDPC code which are trapping sets for various iterative
decoders. It has also been pointed out by several works such as
\cite{Koetter_TC_2003} that the minimum weight of pseudo-codewords is typically
smaller that the minimum distance for most LDPC codes. Thus, the presence of
trapping sets in the Tanner graph of the code in principle prevents the
iterative decoders to approach the performance of MLD for finite lengths LDPC
codes.

An LDPC code ${\cal C}$ is said to have a $t$-guaranteed error correction
capability under a particular decoding algorithm over the Binary Symmetric
channel (BSC) if it can correct all error patterns of weight $t$ or less. The
guaranteed error correction capability of an LDPC code for the BSC plays a
crucial role in its error floor performance as it determines the slope of the
error floor \cite{milos}. Moreover, the problem of guaranteed error correction
is critical for applications such as magnetic, optical and solid-state storage,
flash memories, optical communication over fiber or free-space, as well as an
important open problem in coding theory. Guaranteed error correction is
typically achieved by using Bose-Chaudhuri-Hocquenghem (BCH) or  Reed-Solomon
(RS) codes and hard-decision decoders such as the Berlekamp-Massey decoder
\cite{Book_Coding}, but very little is known about the guaranteed error
correction capability of LDPC codes under iterative decoding. The main reason
for this comes from the fact that even though the error floor performance of an
LDPC code can be relatively well characterized
through the identification of its trapping sets, it is still an arduous task to
determine whether a particular iterative decoder succeeds in correcting all
$t$-error patterns. The guaranteed error correction capability of a particular
LDPC code can vary depending on the particular iterative decoder that is being
used \cite{shashiITW}.

In the first part of our two-part paper series, we introduced a new class of
finite precision iterative decoders, referred to as {\it finite alphabet
iterative decoders} (FAIDs) \cite{Planjery_ElecLetters_2011,shivaisit2010},
which are much lower in complexity than the BP algorithm but can provide a
superior error-rate performance in the error floor region. FAIDs requiring only
a small number of precision bits (as small as three) were shown to surpass BP in
the error floor region on several codes of practical interest due to its ability
to achieve a higher guaranteed error correction capability than the BP algorithm 
\cite{Planjery_ElecLetters_2011,shivaisit2010,Danjean_ITW_2011}. For instance,
on the column-weight-three $(155,64)$ Tanner code, it was shown that
there are $3$-bit precision FAIDs that guarantee a correction of up to $5$
errors, whereas the BP (implemented in floating-point with a maximum of $100$
iterations) fails to correct several $5$-error patterns
\cite{Declercq_ISTC_2010}. 

Despite the superior error floor performance achieved by the FAIDs, their
performance especially in terms of guaranteed error correction capability is
still far from the performance of MLD. For example on the Tanner code, with its
minimum distance $d_{\mrm{min}}=20$, a guaranteed error correction of $5$ errors
achieved by FAIDs is still far from the capability of MLD which is $9$ errors,
therefore leaving large room for improvement. 

In this paper, we
aim at reducing this gap by introducing a general approach that can further
improve the guaranteed error correction capability of LDPC codes. The approach
relies on using a set of carefully chosen FAIDs which are tuned to have
different dynamical behaviors in terms of their error correction on a given
code. The idea is that if an error pattern cannot be corrected by one particular
decoder, there is another decoder in the set that can correct this pattern. The
set of selected FAIDs can then be used (either sequentially or in parallel) to
collectively correct a diverse set of error patterns including some which were
not correctable by a single FAID. This capability of a set of FAIDs to
collectively correct a diverse set of error patterns is referred to as {\it
decoder diversity}. The framework of FAIDs and their simplicity makes them good
candidates for decoder diversity as a plurality of FAIDs can easily be defined
by specifying their variable node update maps.

The main objective of our approach can be summarized as
follows: given a particular LDPC code, we would like to identify a set of FAIDs
that when used sequentially or in parallel can correct a fixed number of errors,
say $t$. A brute force approach would rely on checking all possible error
patterns up to weight $t$ for every FAID considered, and then choosing the set
of FAIDs that correct all the patterns. However, this brute force approach would
be prohibitively complex. Instead, we restrict our attention to only error
patterns associated with the harmful topologies present in the code that could
be trapping sets. Our approach then involves searching for such topologies in
the code, considering all error patterns up to weight $t$ whose support lies in
these topologies, and then finding a combination of FAIDs that can correct all
these particular error patterns. Using the $(155,64)$ Tanner code as an example,
we shall present our methodology in detail and show that the guaranteed error
correction capability of the code can be increased from $t=5$ which is
achievable by using a single FAID to $t=7$ by using decoder diversity.

The rest of the paper is organized as follows. Section \ref{sec:Preliminaries}
provides the necessary preliminaries. Section \ref{sec:diversity} introduces the
concept of decoder diversity and describes our general approach. In Section
\ref{sec:casestudy}, we use the $(155,64)$ Tanner code as a case study and
discuss in detail how our approach can be used to increase the guaranteed error
correction capability of the code. Finally, conclusions are presented in Section
\ref{sec:conclusions}.  
\section{Preliminaries}
\label{sec:Preliminaries}

The Tanner graph $G$ of an $(N,K)$ binary LDPC code $\mathcal{C}$ is a bipartite
graph with two sets of nodes: the set of variable nodes
$V=\{v_1,\cdots,v_N\}$ and the set of check nodes $C=\{c_1,\cdots,c_M\}$. The
set of neighbors of a node $v_i$ is denoted as $\mathcal{N}(v_i)$, and the set
of neighbors of node $c_j$ is denoted by $\mathcal{N}(c_j)$. The degree of a node
is the number of its neighbors. We shall consider only LDPC codes with regular
column-weight $d_v$, where all variable nodes have the same degree $d_v$.

Let $\mathbf{x}=(x_1,x_2,\ldots,x_N)$ denote a codeword of $\cal C$ that is
transmitted over the BSC, where $x_i$ denotes the value of the bit associated
with variable node $v_i$, and let the channel output vector be denoted as
$\mathbf{r}=\{r_1,r_2,\ldots,r_N\}$. Let $\mathbf{e}=(e_1,e_2,\ldots,e_N)$ be
the \emph{error pattern} introduced by the BSC such that $\mathbf{r} =
\mathbf{x} \oplus \mathbf{e}$, and $\oplus$ is the modulo-two sum operator. The
support of an error pattern $\mathbf{e}$, denoted by $\supp(\mathbf{e})$, is
defined as the set of all positions $i$ such that $e_i\neq 0$. The \emph{weight}
of the error pattern $\mathbf{e}$, denoted by $w(\mathbf{e})$ is the cardinality
of $\supp(\mathbf{e})$. 
Let $\mathbf{y}=(y_1,y_2,\ldots,y_N)$ denote the input vector to the decoder,
where each $y_i$ also referred to as a {\it channel value}, is calculated at a
node $v_i$ based on the received value $r_i$.

\subsection{Finite alphabet iterative decoders}
An $N_s$-level FAID denoted by ${\mathrm D}$, is a 4-tuple given by ${\mathrm
D}=(\mathcal{M},\mathcal{Y},\Phi_v,\Phi_c)$. The messages are levels confined to
a finite alphabet $\mathcal M = \{-L_s,\ldots,-L_2,-L_1,0,L_1,L_2,\ldots,L_s\}$
consisting of $N_s=2s+1$ levels, where $L_i\in\mathbb{R^{+}}$ and $L_i>L_j$ for
any $i>j$. 
The sign of a message $x \in \mathcal M$ can be interpreted as the estimate of
the bit (positive for zero and negative for one) associated with the variable
node for which $x$ is being passed to (or from), and the magnitude $|x|$ as a
measure of how reliable this value is. The message $0$ in the alphabet can be
interpreted as an erasure message. 

The set $\mathcal{Y}$ denotes the set of possible channel values. For the case
of BSC, $\mathcal{Y}=\{\pm \mathrm{C}\}$, where $\mathrm{C}\in\mathbb{R^{+}}$.
By convention, we use the mapping $0\rightarrow \mathrm{C}$ and $1\rightarrow -\mathrm{C}$. 
Let $m_1,m_2,\ldots,m_{l-1}$ denote the $l-1$ extrinsic incoming messages of a
node (check or variable) of degree $l$ which are used in the calculation of the
outgoing message.

The function $\Phi_c: \mathcal{M}^{d_c-1} \to \mathcal{M} $ is used for update
at a check node with degree $d_c$ and is defined as
\begin{equation}
\label{eq:phic}
\Phi_c(m_1,\ldots,m_{d_c-1}) = \left(\prod_{j=1}^{d_c-1}\sgn(m_j)\right) \min_{j
\in \{1,\ldots,d_c-1\}}(|m_j|)
\end{equation}

The function $\Phi_v:\mathcal{Y} \times \mathcal{M}^{d_v-1} \to \mathcal{M} $ is
a map used for update at a variable node with degree $d_v$. 

It can described as a closed form function or simply as a $d_{v-1}$-dimensional array or
look-up table (LUT). More details on the closed-form description are provided in
the first part of our two-part series of papers. In this paper, we shall
only use the LUT form which is convenient for defining multiple update maps
required for decoder diversity. 

Note that the maps defining $\Phi_v$ must satisfy the symmetry property which is
\begin{equation}
\Phi_v(y_i,m_1,\ldots,m_{d_v-1}) = - \Phi_v(-y_i,-m_1,\ldots,-m_{d_v-1})
\label{eq:symm}
\end{equation}
and the lexicographic ordering property which is
\begin{equation}
\label{eq:lexicographic}
\Phi_v(\mrm{-C},m_1,\ldots,m_{d_v-1}) \geq
\Phi_v(\mrm{-C},m'_1,\ldots,m'_{d_v-1})
\end{equation} 

$\forall$ $i\in\{1,\ldots,d_v-1\}$ such that $m_i \geq m'_i$ 

Let us alternatively define $\mathcal{M}$ to be $\mathcal{M}=\{M_1,M_2,\cdots,
M_{N_s}\}$ where $M_1=-L_s$, $M_2=-L_{s-1}$,$\cdots$,
$M_s=-L_1$, $M_{s+1}=0$, $M_{s+2}=L_2$,$\cdots$, $M_{N_s}=L_s$. For
column-weight $d_v=3$ codes, the function $\Phi_v$ can be conveniently
represented as a two-dimensional array $[l_{i,j}]_{1 \leq i\leq N_s,1\leq j \leq
N_s}$, where $l_{i,j} \in \mathcal M$, such that
$\Phi_v(M_i,M_j,\mrm{-C})=l_{i,j}$
for any $M_i,M_j\in\mathcal{M}$. The values for $\Phi_v(M_i,M_j,\mrm{+C})$ can
be deduced from the symmetry of $\Phi_v$.
The notations used for the LUT representation of $\Phi_v$ for a 7-level FAID are
shown in Table \ref{tab:LUTdata}, and some examples are listed in Appendix \ref{sec:FAIDrules}. 

\subsection{Trapping sets}
\label{sec:TrappingSets}
For a given decoder input $\mathbf{y}=\{y_1,y_2,\cdots,y_N\}$, a trapping set
(TS) $\mathbf{T}(\mathbf{y})$ is a non-empty set of variable nodes that are not
eventually corrected by the iterative decoder \cite{richardsonerrorfloors}. A
standard notation commonly used to denote a trapping set is $(a,b)$, where $a=|\mathbf{T}(\mathbf{y})|$, and $b$ is the
number of odd-degree check nodes present in the subgraph induced by $\mathbf{T}(\mathbf{y})$.

The Tanner graph representation of an $(a,b)$ TS denoted by $\mathcal{T}$ is a
subgraph induced by $\mathbf{T}(\mathbf{y})$ containing $a$ variable nodes and
$b$ odd-degree check nodes. A code $\mathcal{C}$ is said to contain a TS of type
$\mathcal{T}$ if there exists a set of variable nodes in $G$ whose induced
subgraph is isomorphic to $\mathcal{T}$, seen as a topological structure. Let
$N_\mathcal{T}$ denote the number of trapping sets of type $\mathcal{T}$ that
are contained
in the code $\mathcal{C}$. Also for convenience we shall simply use $\mathbf{T}$
(instead of the more precise notation $\mathbf{T}(\mathbf{y})$) to refer to a
particular subset of variable nodes in a given code that form a trapping set.
Finally, let $\{\mathbf{T}_{i,\mathcal{T}}\ | \ i=1,\ldots,N_{\mathcal{T}}\}$ be
the collection of trapping sets of type $\mathcal{T}$ present in code
$\mathcal{C}$. In other words, $\{\mathbf{T}_{i,\mathcal{T}}\}_i$ is a
collection of all
distinct subsets of variable nodes whose induced subgraphs are isomorphic to
trapping sets of type $\mathcal{T}$.

A TS is said to be elementary if $\mathcal{T}$ contains only degree-one or/and
degree-two check nodes. It is well known that the error floor phenomenon is
dominated by the presence of elementary trapping sets \cite{richardsonerrorfloors,shashierrorfloor}. Hence, throughout this
paper, we shall only consider elementary trapping sets.

Although the $(a,b)$ notation is typically used in literature, this notation is
not sufficient to uniquely denote a particular trapping set as there can be many
trapping sets with different topological structures that share the same values
of $a$ and $b$. This is important to consider since the topological structure of
a particular $(a,b)$ TS determines how harmful the TS is for the error floor of
a given decoder \cite{ontology}. On the other hand, a notation which includes
complete topological description of a subgraph would be extremely complicated
and too precise for our purpose. Therefore, we introduce a simplified notation
which only captures the cycle structure of the subgraph thus giving a cycle
inventory of a trapping set.

\begin{definition}
A trapping set is said to be of type $(a,b ; \prod_{k \geq 2} (2k)^{g_k})$ if
the corresponding subgraph contains exactly $g_k$ distinct cycles of length $2k$.
\end{definition}

Our choice of notation appears to be sufficient for differentiating between the
topological structures of multiple $(a,b)$ trapping sets, and also includes the
definition of codewords of $\mathcal{C}$, as the $(a,0)$ trapping sets
corresponds to codewords of weight $a$. 

\section{Decoder Diversity}
\label{sec:diversity}
\subsection{Decoder diversity principle}
We shall now formally introduce the concept of {\em decoder diversity}. Let us
assume that we have at our disposal a set of $N_s$-level FAIDs denoted by
\begin{equation} {\mathcal D} = \left\{
\left(\mathcal{M},\mathcal{Y},\Phi_v^{(i)},\Phi_c\right) \,|\,
i=1,\ldots,N_\mathcal{D} \right\}
\label{eq:decoder_set}
\end{equation}
where each $\Phi_v^{(i)}$ is a uniquely defined map. We refer to this set
$\mathcal{D}$ as a {\it decoder diversity set} with cardinality $N_\mathcal{D}$,
and an element of this set is
denoted by $\mrm{D}_i$ where
$\mrm{D}_i=\left(\mathcal{M},\mathcal{Y},\Phi_v^{(i)},\Phi_c\right)$. 

Given a code $\mathcal{C}$, we would like to determine whether the FAIDs in the
set $\mathcal{D}$ could be used in combination (either sequentially or in
parallel) in order to guarantee the correction of all error patterns up to a
certain weight $t$. We first introduce notations to denote the set of error
patterns correctable by each decoder. 
Let ${\mathcal E}$ denote an arbitrary set of error patterns on a code
$\mathcal{C}$ whose Tanner graph is $G$, {\it i.e.}
a set of vectors $\mathbf{e}$ with non-zero weight. Let ${\mathcal
E}_{\mrm{D}_i} \subseteq {\mathcal E}$ denote the subset of error patterns 
that are correctable by a FAID $\mrm{D}_i\in \mathcal{D}$.
\begin{definition}
We say that the set of error patterns ${\mathcal E}$ is {\it correctable} by a
decoder diversity set ${\mathcal D}$ if
\[ {\mathcal E} = \bigcup_{i=1}^{N_{\mathcal{D}}} \mathcal{E}_{\mrm{D}_i} \]
\end{definition}

Note that at this point, we have not yet placed any limit on the maximum number
of decoding iterations of each decoder $\mrm{D}_i$, and this issue will be
subsequently addressed in Section \ref{sec:casestudy} using the example of the
$(155,64)$ Tanner code. Given a set of error patterns up to a 
certain weight $t$ on the code $\mathcal{C}$, one would like to determine the
smallest decoder diversity set that can correct all such error patterns. This
problem is known as the 
Set Covering Problem, and is NP-hard \cite{karp}. In this paper, we propose a 
greedy algorithm which can provide a decoder diversity set
$\mathcal{D}$ of FAIDs that may not necessarily be the smallest set, but can
still significantly increase the guaranteed error correction capability of a
given code.

\begin{figure}[p]
\centering
\subfigure[]{\includegraphics[width=0.35\columnwidth]{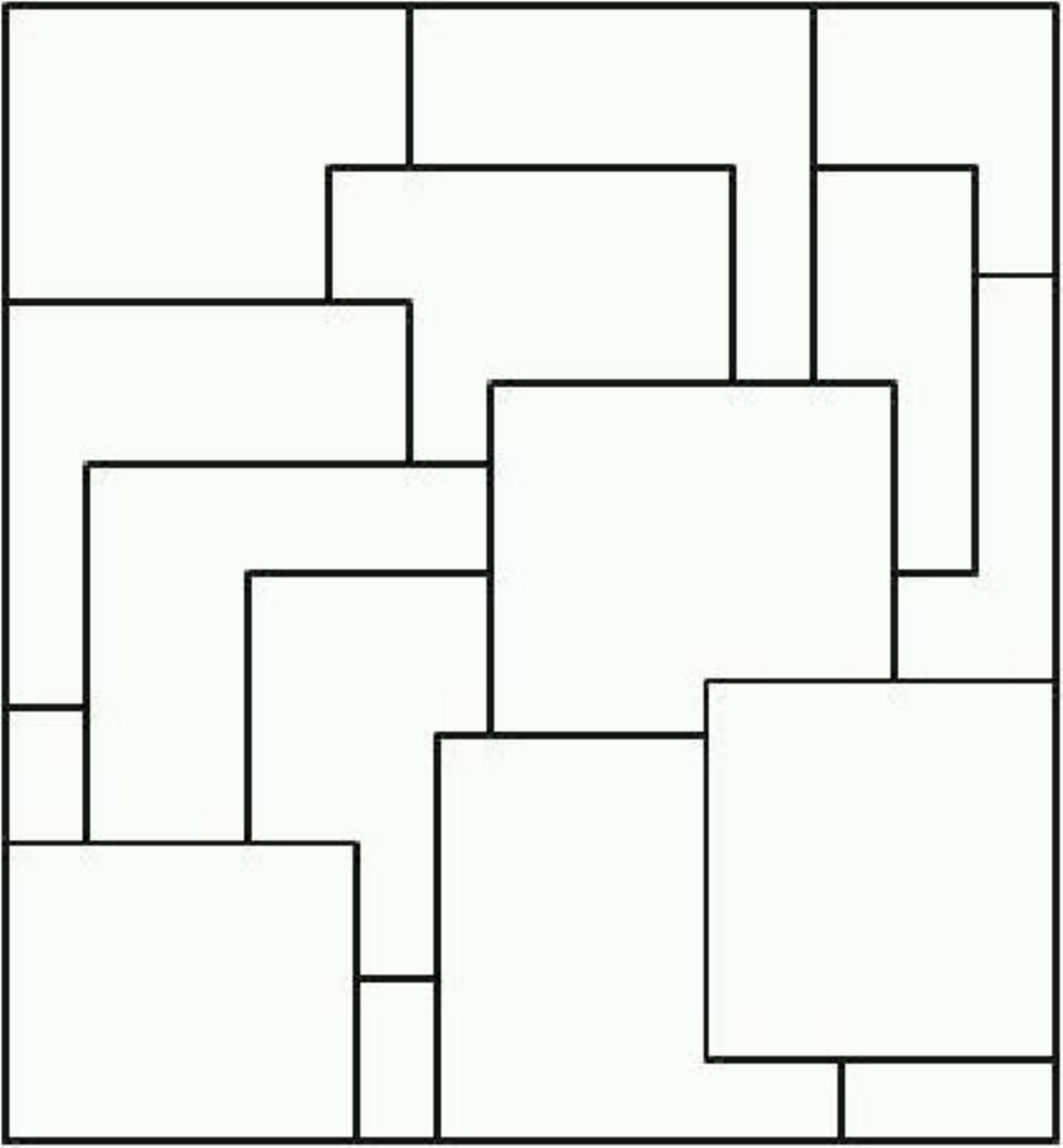}
\label{fig:paving1}}
\hspace*{5mm}
\subfigure[]{\includegraphics[width=0.35\columnwidth]{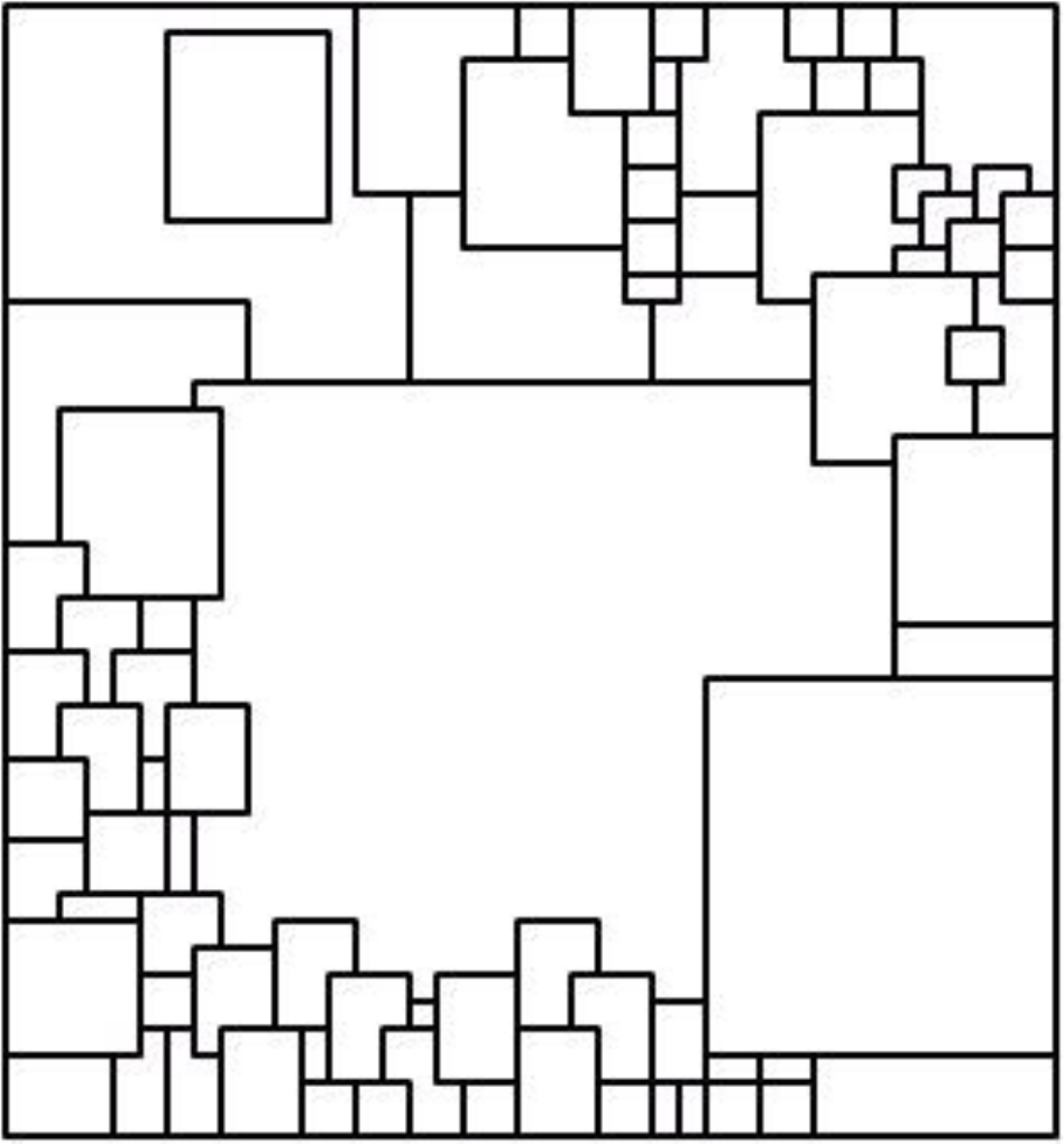}
\label{fig:paving2}}
\caption{Typical ways in which decoder diversity can correct all error patterns
from a pre-determined set $\cal E$.}
\label{fig:paving}
\end{figure}

Note that in the definition of a decoder diversity set, we do not make any a
priori assumptions on the cardinalities of each correctable subset
$\mathcal{E}_{\mrm{D}_i}$. Typically, strong decoders have large correctable
subsets $\mathcal{E}_{\mrm{D}_i}$, while other decoders which are selected to
correct very specific error patterns, could have a small correctable subset.
There are different ways to compose the diversity set ${\mathcal D}$ from
${\mrm{D}_i}$'s in order to cover the set ${\mathcal E}$ with the sets
$\mathcal{E}_{\mrm{D}_i}$. Two distinct ways are illustrated in Fig.
\ref{fig:paving}. Fig. \ref{fig:paving1} shows a case where the set of error
events ${\mathcal E}$ (represented as a big square) is paved with nearly equally
powerful decoders (smaller overlapping squares of similar sizes). Fig.
\ref{fig:paving2} shows another type of covering corresponding to using one
strong decoder and a number of weaker decoders (smaller rectangles) dedicated to
``surgical'' correction of specific error patterns not correctable by the strong
decoder.

\subsection{Error sets}
\label{sec:errorsets}
As mentioned previously, our main goal is to find a, possibly
small, decoder diversity set ${\mathcal D}$ which guarantees correction
of a fixed number of errors $t$. In this section, we describe the 
error sets that will be used for the selection of FAIDs in $\mathcal{D}$.

Let $G'$ be a subgraph that is present in the Tanner graph $G$ of code
$\mathcal{C}$. $G'$ defines typically closed topological structures such as
trapping sets. Let $\mathcal{E}^{k}(G')$ be denoted by the set of all error
patterns 
of weight $k$ whose support lies entirely in the variable node set of subgraph
$G'$:
\begin{equation}
\mathcal{E}^{k}(G')=\{\mathbf{e}: w(\mathbf{e})=k, \ \supp(\mathbf{e})\subseteq
V'\}
\end{equation}
Note that $\mathcal{E}^{k}(G)$ denotes the set of all $k$-error patterns in the code $\mathcal{C}$. 
For simplicity, we shall denote this particular set as $\mathcal{E}^{k}$ instead of $\mathcal{E}^{k}(G)$.
Also let $\mathcal{E}^{[t]}=\bigcup_{k=1}^{t} {\mathcal E^{k}}$ denote the sets of all error patterns whose weight is at most $t$.

A brute force approach to ensure a $t$-guaranteed error correction capability is
to consider all the error patterns in the set $\mathcal{E}^{[t]}$ for the design of the 
diversity set ${\mathcal D}$. Obviously, the cardinality of such an error pattern set is
too large for a practical analysis. 
Instead, we shall consider smaller error pattern sets, based on the knowledge of
the trapping set distribution of the code ${\mathcal C}$. It is reasonable to assume that the errors patterns that
are the most difficult to correct for the iterative decoders are patterns
whose support is concentrated in the topological neighborhood of trapping sets.

Recall that $\{\mathbf{T}_{i,\mathcal{T}}\ | \ i=1,\ldots,N_{\mathcal{T}}\}$
denotes the collection of all $(a,b)$ trapping sets of type $\mathcal{T}$
that are present in code $\mathcal{C}$. Let $\mathcal{E}^k (\mathcal{T})$ denote
the set of error patterns of weight $k$ whose support lies in a $(a,b)$
trapping set $\mathbf{T_{i,\mathcal{T}}}$ of type $\mathcal{T}$. More precisely,
\begin{equation}
\mathcal{E}^k (\mathcal{T})=\{\mathbf{e}: w(\mathbf{e})=k ,
\supp(\mathbf{e})\subseteq \mathbf{T}_{i,\mathcal{T}} \ \ i \in\{1,\ldots,N_{\mathcal{T}}\} \}
\end{equation}

The cardinality of $\mathcal{E}^k (\mathcal{T})$ is given by 
$|\mathcal{E}^k (\mathcal{T})|=\left( \begin{array}{c} a \\ k \end{array}\right)\; N_{\mathcal{T}}$. 

Now, let $\Lambda_{a,b}$ denotes the set of all trapping sets of different types
present in the code $\mathcal{C}$ that have the same parameters $(a,b)$.
The error sets $\mathcal{E}^k (\Lambda_{a,b})$ and $\mathcal{E}^{[t]}
(\Lambda_{a,b})$ associated with $\Lambda_{a,b}$ are defined as follows:
\begin{equation}
\mathcal{E}^k (\Lambda_{a,b}) = \bigcup_{\mathcal{T}\in \Lambda_{a,b}} \mathcal{E}^k (\mathcal{T}) \hspace*{1cm} 
\mathcal{E}^{[t]} (\Lambda_{a,b}) = \bigcup_{k=1}^{t}\mathcal{E}^k(\Lambda_{a,b})
\label{eq:error_sets}
\end{equation}

Finally, $\Lambda^{\{A,B\}}$ is the set containing all $(a,b)$ trapping sets of
different types for different values of $a \leq A$ and $b \leq B$, {\it i.e.} $\Lambda^{(A,B)}=\bigcup_{0 \leq a\leq A, \ 0 \leq b\leq B} \Lambda_{a,b}$ 
and its associated error set is:
\begin{equation}
\mathcal{E}^{[t]} (\Lambda^{(A,B)})=\bigcup_{ 0 \leq a\leq A, \ 0 \leq b\leq
B}\mathcal{E}^{[t]} (\Lambda_{a,b})
\label{eq:error_sets3}
\end{equation}

Clearly, $\mathcal{E}^{[t]} (\Lambda^{(A,B)})\subseteq \mathcal{E}^{[t]}$, and
the cardinality of the latter error set can be further reduced by taking into
account certain structural properties that the Tanner graph of the code may have
due to a specific LDPC code design. 
Quasi-cyclic codes are prime examples of structured codes \cite{Tanner_IT_2004}. 
Tanner graphs of such codes possess many trapping sets that are not only isomorphic 
in the sense of their topological structure, but also have identical neighborhoods. 
Therefore it suffices to consider error patterns associated with any one of these isomorphic 
topologies rather than considering all of them. Certain LDPC code constructions can ensure that the
codes have even more structural properties than just the quasi-cyclicity.
A notable example of constrained algebraic construction is reported in \cite{Tanner_IT_2004}, 
in which the existence of three types of homomorphisms reduces the number of trapping sets of maximum
size $(A,B)$ that need to be considered by several orders of magnitude. 
More details on the example of the $(155,64)$ Tanner code shall be provided in Section \ref{sec:casestudy}.

From the standpoint of computational complexity, it is indeed important to
limit the maximum size of the trapping sets that are included in 
the set $\Lambda^{(A,B)}$. We now provide a conjecture that gives a criterion for the choice
of the values of $A$ and $B$, which are needed for defining the error sets. 

\begin{conjecture}
\label{conj}
If there exists a decoder diversity set $\mathcal{D}$ that corrects all patterns
in the set $\mathcal{E}^{[t]} (\Lambda^{(A,B)})$ on the code $\cal C$ with
$A=2t$ and sufficiently large $B$, then the decoder diversity set $\mathcal{D}$
will also correct all error patterns up to weight $t$ on the code $\mathcal{C}$
with high probability.
\end{conjecture}

This conjecture was found to be valid for the test cases that we have analyzed. The first remark 
concerns the choice of $B$. Typically it has been observed that, in the case of
column-weight $d_v=3$ LDPC codes, most harmful $(a,b)$ trapping sets have small
values of $b$. Note that this is not the case anymore
for LDPC codes with $d_v=4$, as explained with the concept of absorbing sets in
\cite{Dolecek_IT_2010}. 

The above conjecture is analogous to the condition for correcting $t$ errors by
the MLD, which requires the Hamming weight of error patterns to be lower than
$\lfloor d_{min}/2 \rfloor$. In other words, if a decoder $\mrm{D}_i \in
\mathcal{D}$ cannot correct all weight-$t$ error patterns whose support is
entirely contained on trapping sets of size smaller than $2t$, then it is more
likely to not be able to correct more scattered weight-$t$ error patterns as
well: {\it topologically concentrated error patterns are more difficult to correct}.

At the present stage of this work, we have not found any counter-example, but
have not been able to prove the conjecture. We have analyzed several codes, and
for this paper, we present the results of the $(155,64)$ Tanner code for which
the conjecture was verified.

Based on the above conjecture, we now see that considering the set
$\mathcal{E}^{[t]} (\Lambda^{(A,B)})$ instead of $\mathcal{E}^{[t]}$ 
is argued to be sufficient for determining the decoder diversity set that ensures guaranteed
error correction capability of $t$, and this has a significant complexity reduction, 
as will be shown on the $(155,64)$ Tanner code.

\subsection{Generation of FAID diversity sets}
We now present the procedure for obtaining the FAID diversity set that
guarantees the correction of all error patterns in the set $\mathcal{E}^{[t]}
(\Lambda^{(A,B)})$. We shall denote this set by ${\mathcal D}^{[t]}$. 

Let us assume that we are given a large set of candidate FAIDs ${\mathcal D_{base}}$ 
that are considered for possible inclusion into the diversity set.
This set could be obtained from simulations on different codes or by using a
selection technique that was presented in the first part of our two-part series.
Our goal is to build a possibly small set $\mathcal{D}^{[t]}$ from FAIDs
belonging to $\mathcal D_{base}$, that collectively corrects all error patterns
in $\mathcal{E}^{[t]} (\Lambda^{(A,B)})$. In essence, the procedure described 
in algorithm \ref{Decoder Diversity Selection Algorithm} runs over
all error patterns in $\mathcal{E}^{[t]} (\Lambda^{(A,B)})$ and determines their
correctability when decoded by different FAIDs from ${\mathcal D_{base}}$. 
In algorithm \ref{Decoder Diversity Selection Algorithm}, $N_I$ is the maximum number of decoding iterations 
and $\mathcal{E}^r_{\mrm{D}_i}$ denotes the subset of error patterns of $\mathcal{E}^r$ that are correctable by the FAID
$\mrm{D}_i$.

\begin{algorithm}
\caption{Decoder Diversity Selection Algorithm}
\label{Decoder Diversity Selection Algorithm}
\begin{tabular}{cc}
&
\begin{minipage}{0.90\columnwidth}
\small
\begin{enumerate}
\item Given ${\mathcal D_{base}}$ and $N_I$, set ${\mathcal D^{[k]}}=\void$
$\forall (l-1) \leq k \leq t$.\\
Initialize $k=l$, set $(A,B)=(2k,B)$. Set
$\mathcal{E}^r=\mathcal{E}^{k}(\Lambda^{(A,B)})$.
\item set ${\mathcal D^{k}}=\void$ and $i=1$.
\begin{enumerate}
\item If $\mathcal{E}^r=\void$, proceed to Step 3. Else, $\forall \
\mrm{D}_j\in\mathcal{D}_{base}\backslash (\mathcal {D}^{[k-1]}\cup \mathcal
{D}^{k} )$, run each FAID on all error patterns in $\mathcal{E}^r$ for a maximum
of $N_I$ iterations and select the FAID with the largest correctable subset of
error patterns $|\mathcal{E}^r_{\mrm{D}_j}|$, i.e., set
$$\mrm{D}_i=\argmax_{\mrm{D}_j\in\mathcal{D}_{base}\backslash (\mathcal
{D}^{[k-1]}\cup \mathcal {D}^{k} )}|\mathcal{E}^r_{\mrm{D}_j}|.$$Set $\mathcal
D^{k}=\mathcal D^{k} \cup \mrm{D}_i$. 
\item Remove all error patterns corrected by $\mrm{D}_i$ from the set
$\mathcal{E}^r$, i.e., $\mathcal{E}^r=\mathcal{E}^r \backslash
\mathcal{E}^r_{\mrm{D}_i}$.
\item If $\mathcal {E}^r=\void$, proceed to Step 3. Else, proceed to next step. 
\item If $i<|\mathcal{D}_{base}|$, set $i=i+1$ and go back to Step 2a. Else
STOP. The algorithm has failed with the initial parameters of
$\mathcal{D}_{base}$ and $N_I$. 
\end{enumerate}
\item Set ${\mathcal D^{[k]}}={\mathcal D^{[k]}} \cup {\mathcal D^{k}}$.
\item If $k=t$, STOP. The algorithm has successfully built the desired diversity
set $\mathcal{D}^{[t]}$. \\
Else, set $k=k+1$, $(A,B)=(2k,B)$, and ${\mathcal D^{[k]}}={\mathcal
D^{[k-1]}}$.
\begin{enumerate}
\item $\forall$ $\mrm{D}_j \in \mathcal {D}^{[k]}$, determine the correctable
subsets of $k$-error patterns of each FAID $\mrm{D}_j$ denoted by
$\mathcal{E}^{k}_{\mrm{D}_j}(\Lambda^{(A,B)})$.
\item set $\displaystyle \mathcal{E}^r = \mathcal{E}^{k}(\Lambda^{(A,B)})
\backslash \bigcup_{\mrm{D}_j \in {\mathcal D^{[k]}}}
\mathcal{E}^{k}_{\mrm{D}_j}(\Lambda^{(A,B)})$.
\end{enumerate}
\end{enumerate}
\end{minipage}
\end{tabular}
\end{algorithm}

The algorithm starts by building the diversity set $\mathcal{D}^{[k]}$ for a
given $k$, then iteratively expands to the diversity
sets $\mathcal{D}^{[k+1]},\mathcal{D}^{[k+2]}, \ldots, \mathcal{D}^{[t]}$ by
including more and more FAIDs from $\mathcal{D}_{base}$ 
that collectively correct error patterns with increasing weight in $\mathcal{E}^{[t]} (\Lambda^{(A,B)})$. 
The iterative selection of FAIDs is carried out by keeping
track, at each iterative stage, of the set of unresolved error patterns $\mathcal{E}^r \subset
\mathcal{E}^{[t]} (\Lambda^{(A,B)})$ which are not
collectively correctable by the FAIDs selected so far, and then choosing
additional FAIDs to correct these patterns. For example, if 
$\mrm{D}_1,\mrm{D}_2,\ldots,\mrm{D}_L$ are the FAIDs selected so far for
$\mathcal{D}^{[t]}$, and $\mathcal{E}^{[t]}_{\mrm{D}_i}$ denotes the subset of
error patterns correctable by FAID $\mrm{D}_i$, then the set of unresolved error
patterns is 
$$\mathcal{E}^r = \mathcal{E}^{[t]}(\Lambda^{(A,B)}) \backslash \bigcup_{ 1 \leq
i \leq L} \mathcal{E}^{[t]}_{\mrm{D}_i}.$$

The algorithm terminates when  $\mathcal {E}^r=\void$, which means that the set
of FAIDs selected up to that point collectively correct all error patterns in
$\mathcal{E}^{[t]}(\Lambda^{(A,B)})$, and therefore constitute the desired
diversity set ${\mathcal D}^{[t]}$. Assuming that Conjecture \ref{conj} holds,
the obtained diversity set ${\mathcal D}^{[t]}$ will guarantee a correction of
$t$ errors on the LDPC code ${\cal C}$. As a side result, the algorithm also
gives the FAID diversity sets $\mathcal{D}^{[k]}$ for $k<t$.\\

For example, suppose we want to build a decoder diversity set
$\mathcal{D}^{[7]}$ that achieves a guaranteed error correction of $t=7$ on a
code ${\cal C}$, and suppose we know that all FAIDs in the given
$\mathcal{D}_{base}$ guarantee a correction of $t=4$. We then choose an intitial value of $k=5$ 
in Step 1 of the algorithm. The algorithm then starts 
by building the decoder diversity set $\mathcal{D}^{[5]}$ on the
considered error set $\mathcal{E}^{5}(\Lambda^{(10,B)})$ with the given choices
of $\mathcal{D}_{base}$, $N_I$, and $B$. The FAIDs are selected from
$\mathcal{D}_{base}$ in a greedy manner and included in $\mathcal{D}^{[5]}$
until all error patterns in $\mathcal{E}^{5}(\Lambda^{(10,B)})$ are collectively
corrected. Then the algorithm next considers the error set
$\mathcal{E}^{6}(\Lambda^{(12,B)})$ in order to build the set
$\mathcal{D}^{[6]}$. First, all the error patterns correctable by the set
$\mathcal{D}^{[5]}$ are removed from the set $\mathcal{E}^{6}(\Lambda^{(12,B)})$
to constitute the set $\mathcal{E}^{r}$. Then additional FAIDs from
$\mathcal{D}_{base}$ are selected to correct all the error patterns remaining in $\mathcal{E}^{r}$, 
which, together with the FAIDs in $\mathcal{D}^{[5]}$, forms 
the diversity set $\mathcal{D}^{[6]}=\mathcal{D}^{[5]} \cup \mathcal{D}^6$. The
algorithm repeats the procedure for building $\mathcal{D}^{[7]}$ by operating on
the set of error patterns in $\mathcal{E}^{7}(\Lambda^{(14,B)})$. 

Note that the choices of $N_{I}$ and $\mathcal{D}_{base}$ can play an important
role on whether the algorithm is successful or not in building the desired
decoder diversity set. Determining the optimal $N_I$ is beyond the scope of this
paper. However, if the algorithm fails in Step 2d, then increasing the value of
$N_I$ or considering a larger set for $\mathcal{D}_{base}$ typically allows the
algorithm to progress further. We adopted this strategy to obtain a 
decoder diversity set ensuring a $7$-guaranteed error correction on the $(155,64)$ Tanner code, 
as shown in the next section.

\section{Case Study: Guaranteed error correction on the $(155,64)$ Tanner Code}
\label{sec:casestudy}
We shall now use the $(155,64)$ Tanner code \cite{tanner01class,Tanner_IT_2004},
as an example to illustrate how the concept of decoder diversity can be 
used to increase the guaranteed error-correction capability of the code with
reasonable complexity. The $(155,64)$ Tanner code, which is an LDPC code with regular column weight
$d_v=3$ and row weight $d_c=5$, is a particularly good test case for the
following reasons. First, the difference between its minimum distance
$d_{min}=20$ and its minimum pseudo-distance $w_p^{min}\simeq 10$ is large,
which means that the difference in the guaranteed error correction capability
between traditional iterative decoders (Gallager-B, Min-Sum, BP) and the MLD is
expected to be large. Therefore, there is a scope for improvement in reducing
this gap using the approach of FAID decoder diversity. Second, the $(155,64)$
Tanner code is sufficiently small and structured (the code has quasi-cyclicity
equal to 31) so that a brute force checking of whether all error patters up to
certain weight-$t$ are corrected by a decoder diversity set, can be carried out
by Monte Carlo simulations with reasonable computation time.

For comparisons, Table \ref{tab: guaranteed_error_correction_results} shows the
$t$-guaranteed error correction capability of the existing decoders on the
Tanner code. We also found by exhaustively checking through simulations
\cite{Danjean_ITW_2011} that there are no 7-level FAID that can guarantee a
correction of $t>5$ on this particular code. However, using the approach of
decoder diversity, we show that it is possible to increase the guaranteed error
correction capability of the code to $t=7$. As mentioned in the previous section, we only consider error patterns belonging
to $\mathcal{E}^{[t]} (\Lambda^{(A,B)})$ where $A=2t$ and $B$ large enough.
For this code, we verified by simulations that the value of $B=4$ was sufficient
to determine the decoder diversity set ${\mathcal D^{[t]}}$.

The graph structure of this Tanner code satisfies certain properties in
addition to the quasi-cyclicity property \cite{tanner01class}. These properties which are based on
homomorphisms of groups allow for further reduction in the number of error
patterns that need to be considered. Following notations of
\cite{tanner01class}, the transformations $\sigma$, $\pi$, and $\rho$ act on the
indices of the variable nodes and preserve the topological structures. The
transformation $\sigma$ comes from the quasi-cyclicity of the code and allows
then a constant reduction factor of $L=31$ for all the TS topologies, while the
other transformations $\pi$ and $\rho$ can bring another factor of reduction,
depending on the type and location of the TS. More details on the three
different transformations that the Tanner graph of this code follows are
reported in Appendix \ref{sec:homomorphism}. 

The full enumeration of trapping sets with $a \leq 14$ and $b \leq 4$ is
presented in Table \ref{tab:trappingsets}. The first column of the table gives
the $(a,b)$ parameters, and the second column indicates the TS cycle inventory
of different $(a,b)$ TS types (the cycle inventory is omitted for the parameters
that allow too many cycle-inventory types). The last three columns show the
numbers of trapping sets that need to be checked by the Algorithm \ref{Decoder
Diversity Selection Algorithm} when the code homomorphisms are exploited.
$N_\mathcal{T}$ corresponds to the number of trapping sets present without
taking any code structure into account, $N_{\sigma(\mathcal{T})}$ corresponds to
the number of trapping sets present after taking into account the quasi-cyclic
property obtained from the transformation  $\sigma$, and
$N_{\sigma(\pi(\rho(\mathcal{T})))}$ corresponds to the number of trapping sets
present after taking all three transformations $\sigma$, $\pi$ and $\rho$ into
consideration. The small section of the Table at the bottom shows the structure
and number of the lowest weight codewords of different types. 

These trapping sets have been enumerated using the modified impulse algorithm,
which is known as the most efficient algorithm to find low-weight codewords or
near-codewords of a given short length LDPC code
\cite{declercqISIT2008,AbuSurra_ITA_2010}. When the number of types was too
large, we did not indicate the details of the TS notation. It is clear from the
Table that the number of topologies needed to be considered to characterize the
behavior of an iterative decoder on the Tanner code could be greatly reduced.
Actually, the number of structures (including isomorphic) of given type
$\mathcal{T}$ present in the code could be multiples of either
$L\,d_c\,d_v=465$, $L\,d_c=155$ or $L\,d_v=93$ and this number is reduced for
the analysis by the transformations $\sigma(\pi(\rho(\mathcal{T})))$. The TS of
type (5,3;$8^3$) is an example where there are $L\,d_c=155$ such structures in
the Tanner code, while (20,0)-type-III codewords is an example where there are
$L\,d_v=93$ such structures.

\subsection{Error sets for the Tanner code}
The error sets that we have considered for the Tanner code are shown in Table
\ref{tab:error_sets_tanner} along with their cardinalities. The cardinalities of
the error sets have been reduced using the structural properties $\sigma$,
$\pi$, and $\rho$ of the Tanner code to:
\begin{equation}
|\mathcal{E}^k (\mathcal{T})|=\left( \begin{array}{c} a \\ k
\end{array}\right)\; N_{\sigma(\pi(\rho(\mathcal{T})))}.
\end{equation}
where $N_{\sigma(\pi(\rho(\mathcal{T})))}$ is the value obtained from Table
\ref{tab:trappingsets}.

One can further reduce the number of error patterns considered in each error
set, since a particular error pattern belonging to an error set of a small
trapping set may also be included in the error set of a larger trapping set
containing the smaller one. For example, a 5-error pattern on one of the TS
$(9,3)$ could be also listed as one of the 5-error patterns in the TS $(8,2)$ if
$(8,2)$ is contained in $(9,3)$. Therefore, we also take this into account by
including only the error patterns in the error set $\mathcal{E}^{[k]}
(\Lambda_{a,b})$ that are distinct from all error patterns in $\mathcal{E}^{[k]}
(\Lambda_{a',b'})$
with $a'<a$ and $b'<b$. This leads to a further reduction in the number of error
patterns considered in error sets, and the final number is reported at the
bottom of Table \ref{tab:error_sets_tanner}. From the Table, we can see that the
complexity reduction factor in each case is of the order of $10^6$, which is
very large and in any case sufficient to reduce the complexity of finding the
decoder diversity set to a reasonable level.

\subsection{Error correction results for the Tanner code}

Let us recall that we consider only 7-level FAIDs for decoder diversity which
require only 3 bits of precision for their message representation. Our main
results are summarized in the Table \ref{tab: diversity_schemes}. We are able to
guarantee a correction of $t=7$ errors on the Tanner code using
$N_{\mathcal{D}}=343$ FAIDs with $N_{I}=120$ iterations. 

We also verified by brute force Monte Carlo simulations that each of the
obtained diversity sets $\mathcal{D}^{[t]}$ for $t=5,6,7$ guarantees a
correction of all error patterns of weight at most $t$ on the Tanner code even
though only error patterns in $\mathcal{E}^{[t]} (\Lambda^{(A,B)})$ with
$(A,B)=(2t,4)$ were used in the algorithm, thus validating the conjecture stated
in Section \ref{sec:errorsets}.

Due to the huge cardinality reduction in the error sets considered (as shown out
in Table \ref{tab:error_sets_tanner}), we were able to identify the decoder
diversity set for $t=6$ in less than one hour, and for $t=7$ in a few days. Note
that decoder diversity does not require any post-processing, as it is still an
iterative message passing decoder with the additional feature that the the
variable node update rule $\Phi_v$ changes after $N_{I}$ iterations (and the
decoder is restarted). Note also that additional complexity reduction can be
achieved by exploiting any existing similarities between the update rules of the
FAIDs in the decoder diversity set.

In order to illustrate how different 7-level FAIDs in the decoder diversity set
can behave in terms of their correctability of different error patters in the
error set, we provide two examples with the help of Table \ref{tab:statPaving}. 

The first part of the Table \ref{tab:statPaving} shows an example of using
equally {\it powerful} decoders in the diversity set. Statistics are provided on
the number of correctable error patterns $\abs{\mathcal{E}^{[6]}_{\mrm{D}_i}
(\Lambda^{(12,4)})}$ by each 7-level FAID in the decoder diversity set
$\mathcal{D}^{[6]}$ from the error set $\mathcal{E}^{[6]}(\Lambda^{(11,4)})$.
The LUT maps of $\Phi_v$ that define these 7-level FAIDs are reported in Table
\ref{tab:FAIDrules} of Appendix \ref{sec:FAIDrules}. For convenience, we have
noted ${\mathcal D^{[5]}}=\{{\mathbf D_0}\}$ and ${\mathcal D^{[6]}}={\mathcal
D^{[5]}} \cup \{{\mathbf D_1},\ldots,{\mathbf D_8}\}$. Recalling that the total
number of error patterns in $\mathcal{E}^{[6]} (\Lambda^{(11,4)})$ is
$\abs{\mathcal{E}^{[6]} (\Lambda^{(11,4)})}=11829$, we can see that all decoders
in ${\mathcal D^{[6]}}$ are in fact almost equally powerful in terms of the
number of error patterns they correct. However, all $9$ decoders when used in
combination are able to collectively guarantee an error correction of $t=6$. 

The second part of Table \ref{tab:statPaving} provides an example of how certain
decoders, that we refer to as {\it ``surgeon"} decoders, 
can be used to specifically correct certain error patterns not correctable by
the particularly good decoders. The statistics shown in the Table are for eight
different FAIDs (labeled $\mrm{D}_{10}$ to $\mrm{D}_{17}$ for convenience) that
were selected to correct eight particular $7$-error patterns in the error set
$\mathcal{E}^{[7]} (\Lambda_{7,3}) \cup \mathcal{E}^{[7]} (\Lambda_{8,2})$.
These eight different FAIDs are required to separately correct each of these
error patterns. Moreover, in comparison with the statistics obtained from the
FAIDs belonging to ${\mathcal D^{[6]}}$ on the 6-error patterns, these decoders
are not as strong as the first nine decoders ${\mathbf D_{0}}$ to ${\mathbf
D_{8}}$. Five of them especially have very poor behaviors on the 6-error events.

These two examples clearly show that in order to guarantee $t$ error correction,
the decoder diversity sets can pave the error sets in very different manners. In
summary, for $t=6$ error correction, the decoder diversity set behaves roughly
like in Fig. \ref{fig:paving1}, while for $t=7$ error correction,
the decoder diversity set behave more like in Fig. \ref{fig:paving2} using both
powerful and surgeon decoders. The list of FAIDs ${\mathbf D_{0}}$ to ${\mathbf
D_{8}}$ and ${\mathbf D_{10}}$ to ${\mathbf D_{17}}$ are reported in Appendix
\ref{sec:FAIDrules}.

Fig. \ref{fig:Plot_7errors} shows the remaining $7$-error patterns in ${\mathcal
E^{[7]}}$ after the sequential use of the FAIDs in ${\mathcal D^{[5]}}$ followed
by the FAIDs in $\mathcal D^{[6]}\backslash \mathcal{D}^{[5]}$, and then
followed by FAIDs in $\mathcal{D}^{[7]}\backslash \mathcal{D}^{[6]}$.

\begin{figure}[p]
\centering
\includegraphics[width=0.8\columnwidth]{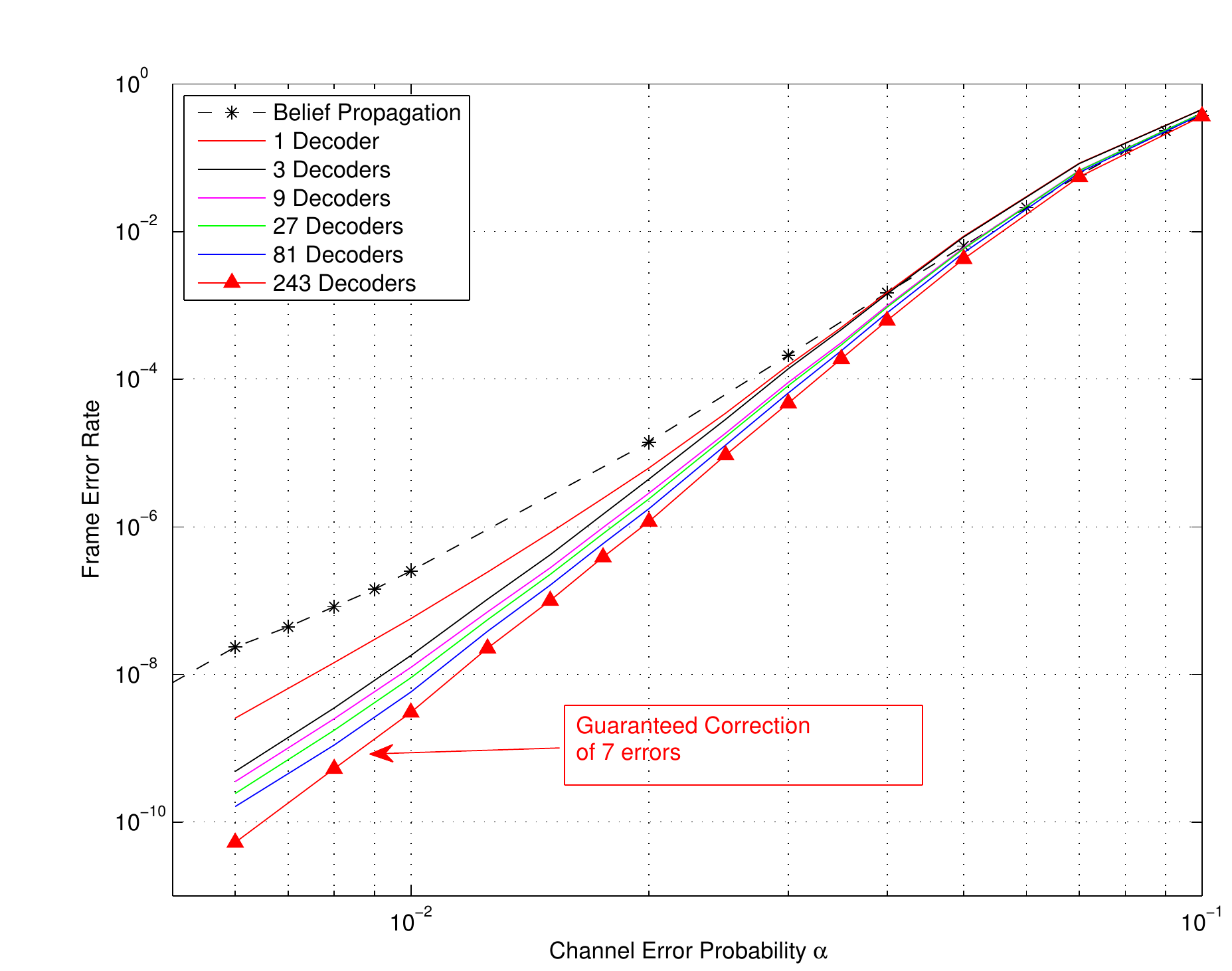}
\caption{FER results on the Tanner Code with guaranteed error correction of 7
errors.}
\label{fig:FER_Tanner_7errors}
\end{figure}

Fig. \ref{fig:FER_Tanner_7errors} shows the FER performance of the decoder
diversity set ${\mathcal D^{[7]}}$, when simulated on the Tanner code
over the BSC channel with cross-over error probability $\alpha$ and with a
maximum of $N_{I}=120$ decoding iterations for each decoder.
One can see that, especially in the error floor region, the use of an increasing
number of FAIDs increases the slope of the FER curve, and eventually reaches 
a slope of $t=8$, which corresponds to the minimum weight error pattern that is
not corrected by our decoder diversity set $\mathcal{D}^{[7]}$.

\section{Conclusions}
\label{sec:conclusions}
We introduced a general decoding scheme that utilizes a collection of several
different 
FAIDs, which is referred to as a decoder diversity set, in order to further
increase the guaranteed error correction capability of a given LDPC code from
what is achievable by a single FAID. We provided a methodology to build the
decoder diversity sets based on using the trapping set distribution of the code,
and considering error patterns that are only associated with the trapping sets
present in the code. Using the $(155,64)$ Tanner code as an example, we showed
that the structural properties of the code can be exploited to reduce the
complexity in terms of reducing the number of considered error patterns by
several orders. We were able to increase the guaranteed error correction
capability of the $(155,64)$ Tanner code using our approach of decoder
diversity, from 
$t=5$ errors that is achievable by a single FAID to $t=7$ errors by FAID decoder
diversity. Note that the BP algorithm is able to guarantee a correction of only
$t=4$ on the Tanner code. Although our discussion throughout this paper
primarily focused on the particular example of the Tanner code, the technique
can be applied to other codes with regular column weight $d_v$, provided that
the Trapping set distribution is known (which is a reasonable assumption for
short to moderate codeword lengths).

\appendices
\section{Topologies of the Tanner Code}
\label{sec:homomorphism}
As explained in \cite{tanner01class}, there are three types of homomorphisms
which preserve the topological structures 
in the graph of the Tanner code, due to the fact that Tanner's design of the
parity-check matrix is based on an array
of $(d_v,d_c)$ circulants of size $L$, and that the values of shifts for the
circulant matrices are chosen from two multiplicative sub-groups
of the Galois field GF$(L)$. For easy understanding, we shall instead present
the homomorphisms as simple transformations acting on the indices of the
variable nodes in the code. Let $\alpha$ (respectively $\beta$) be two elements
of GF$(L)$ with multiplicative order $d_c$ (respectively $d_v$).
The parity check matrix is defined by an array of circulants with shift orders
$\{ \alpha^t \beta^r \}_{0\leq t \leq d_c-1, 0\leq r \leq d_v-1}$.
Now, let the index of a variable node $v_i$ be expressed as $i=k*L+l$. We now
define the three following transformations acting on the indices of $\mathbf{T}$
that preserve the topology as well as the neighborhood of the TS.
\begin{itemize}
\item block-cyclicity: Let $\sigma:\ V\times\{0,\ldots,L-1\}\rightarrow V$. Then
$$\sigma(v_i,t) = v_j \ \ \mrm{where} \ j=(k*L)+(l+t\mod(L))$$

\item row-wise transformation: Let $\pi: \ V\times\{0,\ldots,d_c-1\}\rightarrow
V$. Then
$$\pi(v_i,t) = v_j \ \ \mrm{where} \ j=((k+t\mod(d_c))*L)+(\alpha^t l\mod(L))$$

\item column-wise homomorphism: Let $\rho: \
V\times\{0,\ldots,d_v-1\}\rightarrow V$. Then
$$\rho(v_i,t) = v_j \ \ \mrm{where} \ j=(k*L)+(\beta^t l\mod(L))$$
\end{itemize}

Consider a trapping set of size $a$ bits denoted by $\mathbf{T} = \{ v_{n_1},
\ldots, v_{n_a}\}$. By applying the transformation $\sigma$ on $\mathbf{T}$
such that $\sigma(\mathbf{T},t) = \{ \sigma(v_{n_1},t), \ldots,
\sigma(v_{n_a},t)\}$ where $t\in \{0,\ldots,L-1\}$, the induced subgraphs of
$\sigma(\mathbf{T},t)$
and $\mathbf{T}$ are {\it isomorphic} to each other in the code $\forall t\in
\{0,\ldots,L-1\}$, i.e., they have exactly the same topology and neighborhood.
This implies that one has to only consider error patterns associated with one of
the isomorphic structures instead of all of them. The same applies for
the transformations $\pi$ and $\rho$. By applying all three transformations
$\sigma(\pi(\rho(\mathbf{T})))$, the number of trapping sets of a certain type
$\mathcal{T}$ that need to be considered is significantly reduced.

\section{List of 7-level FAIDs used}
\label{sec:FAIDrules}
In the Table \ref{tab:FAIDrules}, we list some of the 7-level FAIDs used in this
paper, and which were used in  
the FAID diversity sets described in Section \ref{sec:casestudy}. We only
indicate the entries of the LUT array (see Table \ref{tab:LUTdata}) that cannot
be 
deduced by symmetry.


\newpage

\begin{table}[htbp]
	\caption{lut representation of $\Phi_v(\mrm{-C},m_1,m_2)$ for a 7-level
faid}
	\label{tab:LUTdata}
	\centering
\renewcommand{\arraystretch}{1.5}
\resizebox{9cm}{!}{
	\begin{tabular}{|c||c|c|c|c|c|c|c|}
	\hline
		\boldmath $m_{1}/m_{2}$   	& \boldmath$-L_3$
&\boldmath$-L_2$	& \boldmath$-L_1$ 	& \boldmath $0$ 	&
\boldmath$+L_1$ 	& \boldmath $+L_2$ & \boldmath $+L_3$\\ \hline\hline
		\boldmath$-L_3$ 		& $l_{1,1}$	& $l_{1,2}$
& $l_{1,3}$  	&  $l_{1,4}$	& $l_{1,5}$	&  $l_{1,6}$    &  $l_{1,7}$  \\
\hline
		\boldmath$-L_2$ 		& $l_{2,1}$	& $l_{2,2}$ 
& $l_{2,3}$  	&  $l_{2,4}$	& $l_{2,5}$	&  $l_{2,6}$    &  $l_{2,7}$  \\
\hline
		\boldmath$-L_1$			& $l_{3,1}$	& $l_{3,2}$  
& $l_{3,3}$ 	&  $l_{3,4}$	& $l_{3,5}$	&  $l_{3,6}$	&  $l_{3,7}$  \\
\hline
		\boldmath $0$			& $l_{4,1}$	& $l_{4,2}$ 
& $l_{4,3}$  	&  $l_{4,4}$	& $l_{4,5}$	&  $l_{4,6}$	&  $l_{4,7}$  \\
\hline
		\boldmath $+L_1$		& $l_{5,1}$	& $l_{5,2}$ 
& $l_{5,3}$ 	&  $l_{5,4}$ 	& $l_{5,5}$	&  $l_{5,6}$	&  $l_{5,7}$  \\
\hline
		\boldmath $+L_2$		& $l_{6,1}$	& $l_{6,2}$  
& $l_{6,3}$ 	&  $l_{6,4}$ 	& $l_{6,5}$	&  $l_{6,6}$	&  $l_{6,7}$  \\
\hline
		\boldmath $+L_3$		& $l_{7,1}$	& $l_{7,2}$  
& $l_{7,3}$  	&  $l_{7,4}$ 	& $l_{7,5}$	&  $l_{7,6}$	&  $l_{7,7}$  \\
\hline
	\end{tabular}
}
\end{table}

\begin{table}[p]
\caption{\small $t$-guaranteed error correction capability of different decoders
on the $(155,64)$ tanner code}
\label{tab: guaranteed_error_correction_results}
\centering
\begin{tabular}{|l|r|r|} \hline
$t$ & Algorithm & Reference \\ \hline
$3$ & Gallager A and Gallager B & \cite{shashiITW} \\ \hline
$4$ & Min-Sum and Belief Propagation & \cite{shivaisit2010} \\ \hline
$5$ & 5-level and 7-level FAIDs & \cite{Declercq_ISTC_2010}\\ \hline
\end{tabular}
\end{table}

\begin{table}[p]
\caption{\small trapping set spectrum and low-weight codewords spectrum of the
$(155,64)$ tanner Code}
\label{tab:trappingsets}
\renewcommand{\arraystretch}{1.2}
\centering
\resizebox{0.5\columnwidth}{!}{
\begin{tabular}{|l|l|c|c|c|} \hline
$\mathcal{T}$ & TS-label & $N_{\mathcal{T}}$ & $N_{\sigma(\mathcal{T})}$ &
$N_{\sigma(\pi(\rho(\mathcal{T})))}$ \\ \hline
(5,3) & (5,3;$8^3$) & {\bf 155} & {\bf 15} & {\bf 1} \\ \hline
(6,4) & (6,4;$8^1 10^2$) & {\bf 930} & {\bf 30} & {\bf 2} \\ \hline
(7,3) & (7,3;$8^3 10^2 14^2$) & {\bf 930} & {\bf 30} & {\bf 2} \\ \hline
(8,2) & (8,2;$8^3 10^4 12^2 14^4 16^2$) & {\bf 465} & {\bf 15} & {\bf 1} \\
\hline
(8,4) & 4 types & {\bf 5012} & {\bf 165} & {\bf 11} \\ \cline{2-5}
      & (8,4;$8^3 12^2 16^2$) & {} & {45} & {3} \\
      & (8,4;$8^1 10^2 12^2 14^2$) & {} & {15} & {1} \\
      & (8,4;$8^1 10^3 12^1 14^1 16^1$) & {} & {90} & {6} \\
      & (8,4;$8^1 10^4 16^2$) & {} & {15} & {1} \\ \hline
(9,3) & 3 types & {\bf 1860} & {\bf 60} & {\bf 4} \\ \cline{2-5}
      & (9,3;$8^1 10^4 12^4 14^2 16^2 18^2$) & {} & {15} & {1} \\
      & (9,3;$8^1 10^5 12^2 14^2 16^4$) & {} & {30} & {2} \\
      & (9,3;$8^3 10^2 12^2 14^4 16^2 18^2$) & {} & {15} & {1} \\ \hline
(10,2) & 2 types & {\bf 1395} & {\bf 45} & {\bf 3} \\ \cline{2-5}
       & (10,2;$8^1 10^6 12^5 14^4 16^6 18^5 20^2$) & {} & {15} & {1} \\
       & (10,2;$8^3 10^5 12^2 14^6 16^6 18^2 20^4$) & {} & {30} & {2} \\ \hline
(10,4) & 27 types & {\bf 29295} & {\bf 945} & {\bf 63} \\ \hline

(11,3) & 11 types & {\bf 6200} & {\bf 200} & {\bf 14} \\ \hline
(12,2) & 2 types & {\bf 930} & {\bf 30} & {\bf 2} \\ \cline{2-5}
       & (12,2;$8^1 10^6 12^6 14^6 16^9 18^9 20^8 22^7 24^4$) & {} & {15} & {1}
\\
       & (12,2;$8^4 10^2 12^4 14^4 16^8 18^12 20^6 22^8 24^2$) & {} & {15} & {1}
\\ \hline
(12,4) & 170 types & {\bf 196440} & {\bf 6240} & {\bf 416} \\ \hline
(13,3) & 53 types & {\bf 34634} & {\bf 1155} & {\bf 79} \\ \hline
\end{tabular}}
\vspace*{3mm}

\resizebox{0.3\columnwidth}{!}{
\begin{tabular}{|l|l|l|l|l|} \hline
$\mathcal{T}$ & TS-label & $N_{\mathcal{T}}$ & $N_{\sigma(\mathcal{T})}$ &
$N_{\sigma(\pi(\rho(\mathcal{T})))}$ \\ \hline
(20,0) & 3 types & {\bf 1023} & {\bf 33} & {\bf 3} \\ \cline{2-5}
       & type-I & {465} & {15} & {1} \\
       & type-II & {465} & {15} & {1} \\
       & type-III & {93} & {3} & {1} \\ \hline
(22,0) & 14 types & {\bf 6200} & {\bf 200} & {\bf 14} \\ \hline
(24,0) & 97 types & {\bf 43865} & {\bf 1415} & {\bf 97} \\ \hline
\end{tabular}}
\end{table}

\begin{table}[p]
\caption{cardinalities of error sets considered for the $(155,64)$ tanner code.}
\label{tab:error_sets_tanner}
\renewcommand{\arraystretch}{1.2}
\centering
\resizebox{0.9\columnwidth}{!}{
\begin{tabular}{|l|l|l|l|l|l|l|l|l|} \hline
\multicolumn{3}{|c|}{5-errors} & \multicolumn{3}{|c|}{6-errors} &
\multicolumn{3}{|c|}{7-errors} \\ \cline{1-3}
$\mathcal{E}^{[5]} (\Lambda_{5,3})$ & 1 & 1 &  \multicolumn{3}{|c|}{} & 
\multicolumn{3}{|c|}{} \\ \cline{1-6}
$\mathcal{E}^{[5]} (\Lambda_{6,4})$ & 12 & 12 & $\mathcal{E}^{[6]}
(\Lambda_{6,4})$ & 2 & 2 &   \multicolumn{3}{|c|}{}   \\ \hline
$\mathcal{E}^{[5]} (\Lambda_{7,3})$ & 42 & 23 & $\mathcal{E}^{[6]}
(\Lambda_{7,3})$ & 14 & 11 & $\mathcal{E}^{[7]} (\Lambda_{7,3})$ & 2 & 2 \\
\hline
$\mathcal{E}^{[5]} (\Lambda_{8,2})$ & 56 & 20 & $\mathcal{E}^{[6]}
(\Lambda_{8,2})$ & 28 & 15 & $\mathcal{E}^{[7]} (\Lambda_{8,2})$ & 8 & 6 \\
\hline
$\mathcal{E}^{[5]} (\Lambda_{8,4})$ & 616 & 398 & $\mathcal{E}^{[6]}
(\Lambda_{8,4})$ & 308 & 240 & $\mathcal{E}^{[7]} (\Lambda_{8,4})$ & 88 & 79 \\
\hline
$\mathcal{E}^{[5]} (\Lambda_{9,3})$ & 504 & 100 & $\mathcal{E}^{[6]}
(\Lambda_{9,3})$ & 336 & 110 & $\mathcal{E}^{[7]} (\Lambda_{9,3})$ & 144 & 72 \\
\hline
$\mathcal{E}^{[5]} (\Lambda_{10,2})$ & 756 & 399 & $\mathcal{E}^{[6]}
(\Lambda_{10,2})$ & 630 & 416 & $\mathcal{E}^{[7]} (\Lambda_{10,2})$ & 360 & 277
\\ \hline
$\mathcal{E}^{[5]} (\Lambda_{10,4})$ & 15\,876 & 7\,064 & $\mathcal{E}^{[6]}
(\Lambda_{10,4})$ & 13230 & 7860 & $\mathcal{E}^{[7]} (\Lambda_{10,4})$ & 7560 &
5421 \\ \hline
\multicolumn{3}{c|}{}                		    & $\mathcal{E}^{[6]}
(\Lambda_{11,3})$ & 6468 &  1958 & $\mathcal{E}^{[7]} (\Lambda_{11,3})$ & 4620 &
1894 \\ \cline{4-9}
\multicolumn{3}{c|}{}                		    & $\mathcal{E}^{[6]}
(\Lambda_{12,2})$ & 1848 & 766 & $\mathcal{E}^{[7]} (\Lambda_{12,2})$ & 1584 &
857 \\ \cline{4-9}
\multicolumn{3}{c|}{}                		    & $\mathcal{E}^{[6]}
(\Lambda_{12,4})$ & 384\,384 & 163\,562 & $\mathcal{E}^{[7]} (\Lambda_{12,4})$ &
329\,472 & 187\,360 \\ \cline{4-9}
\multicolumn{6}{c|}{}                                                      &
$\mathcal{E}^{[7]} (\Lambda_{13,3})$ & 135\,564 & 31\,890 \\ \cline{7-9}
\multicolumn{6}{c|}{}                                                      &
$\mathcal{E}^{[7]} (\Lambda_{14,2})$ & 37\,752 & 8\,157 \\ \cline{7-9}
\multicolumn{6}{c|}{}                                                      &
$\mathcal{E}^{[7]} (\Lambda_{14,4})$ & 9\,129\,120 & 3\,326\,862 \\ \cline{7-9}
\end{tabular}}
\vspace*{3mm}

\resizebox{0.9\columnwidth}{!}{
\begin{tabular}{|l|l|l|l|l|l|} \hline
\multicolumn{2}{|c|}{5-errors} & \multicolumn{2}{|c|}{6-errors} &
\multicolumn{2}{|c|}{7-errors} \\ \hline
$\mathcal{E}^{[5]} (\Lambda^{(10,4)})$ & 8\,017 & $\mathcal{E}^{[6]}
(\Lambda^{(12,4)})$  & 174\,940 & $\mathcal{E}^{[7]} (\Lambda^{(14,4)})$  &
3\,562\,877 \\ \hline
${\mathcal E^{[5]}}$ & 698\,526\,906 & ${\mathcal E^{[6]}}$ & 17\,463\,172\,650
& ${\mathcal E^{[7]}}$ &  371\,716\,103\,550 \\ \hline
\multicolumn{2}{|c|}{Comp. Reduction Factor} & \multicolumn{2}{|c|}{Comp.
Reduction Factor} & \multicolumn{2}{|c|}{Comp. Reduction Factor} \\ \hline
            \multicolumn{2}{|c|}{87\,130}    &     \multicolumn{2}{|c|}{99\,824}
    &      \multicolumn{2}{|c|}{104\,330}   \\ \hline
\end{tabular}}
\end{table}

\begin{table}[p]
\caption{\small }
\label{tab: diversity_schemes}
\centering
\begin{tabular}{|c|l|l|}  \hline 
$t$ & $N_{\mathcal{D}}$ & $N_{I}$ \\ \hline
$5$ & 1 & 15  \\ \hline
$6$ & 9 & 50  \\ \hline
$7$ & 243 & 120  \\ \hline
\end{tabular}
\end{table}

\begin{table}[p]
\caption{statistics on the error correction of several faids used in the decoder
diversity sets.}
\label{tab:statPaving}
\renewcommand{\arraystretch}{1.2}
\begin{center}
\resizebox{0.5\columnwidth}{!}{
\begin{tabular}{r|c|c|c|c|c|c|c|c|c|} \cline{2-10}
Decoder $\mrm{D}_i$ & $\mrm{D}_0$ & $\mrm{D}_1$ & $\mrm{D}_2$ & $\mrm{D}_3$ &
$\mrm{D}_4$ & $\mrm{D}_5$ & $\mrm{D}_6$ & $\mrm{D}_7$ & $\mrm{D}_8$ \\ \hline
$\abs{\mathcal{E}^{[6]}_{\mrm{D}_i} (\Lambda^{(12,4)})}$ & 11724 & 11779 & 11777
& 11782 & 11784 & 11770 & 11759 & 11623 & 11724 \\ \hline
remaining errors & 105 & 16 & 10 & 7 & 4 & 3 & 2 & 1 & 0 \\ \cline{2-10}
\end{tabular}}
\vspace*{3mm}

\resizebox{0.5\columnwidth}{!}{
\begin{tabular}{r|c|c|c|c|c|c|c|c|} \cline{2-9}
Decoder $\mrm{D}_i$ & $\mrm{D}_{10}$ & $\mrm{D}_{11}$ & $\mrm{D}_{12}$ &
$\mrm{D}_{13}$ & $\mrm{D}_{14}$ & $\mrm{D}_{15}$ & $\mrm{D}_{16}$ &
$\mrm{D}_{17}$\\ \hline
$\abs{\mathcal{E}^{[7]}_{\mathcal{D}_i} (\Lambda_{7,3}) \cup
\mathcal{E}^{[7]}_{\mathcal{D}_i} (\Lambda_{8,2})}$ & 1 & 1 & 1 & 1 & 1 & 1 & 1
& 1 \\ \hline
$\abs{\mathcal{E}^{[6]}_{\mathcal{D}_i} (\Lambda^{(12,4)})}$ & 10781 & 110 & 83
& 208 & 10143 & 3726 & 164 & 321 \\ \cline{2-9}
\end{tabular}}
\end{center}
\end{table}

\begin{table}[p]
\caption{list of some 7-level FAIDs used in this paper. the first nine FAIDs
guarantee an error correction of $t=6$ on the $(155,64)$ Tanner Code.}
\label{tab:FAIDrules}
\renewcommand{\arraystretch}{1.2}
\begin{center}
\resizebox{\columnwidth}{!}{
\begin{tabular}{
|c|c|c|c|c|c|c|c||c|c|c|c|c|c||c|c|c|c|c||c|c|c|c||c|c|c||c|c||c|} \hline
FAID & $l_{1,1}$ & $l_{1,2}$ & $l_{1,3}$ & $l_{1,4}$ & $l_{1,5}$ & $l_{1,6}$ &
$l_{1,7}$ &
$l_{2,2}$ & $l_{2,3}$ & $l_{2,4}$ & $l_{2,5}$ & $l_{2,6}$ & $l_{2,7}$ &
$l_{3,3}$ & $l_{3,4}$ & $l_{3,5}$ & $l_{3,6}$ & $l_{3,7}$ &
$l_{4,4}$ & $l_{4,5}$ & $l_{4,6}$ & $l_{4,7}$ &
$l_{5,5}$ & $l_{5,6}$ & $l_{5,7}$ &
$l_{6,6}$ & $l_{6,7}$ &
$l_{7,7}$ \\ \hline \hline
${\mathbf D_{0}}$ & $-L_3$ & $-L_3$ & $-L_3$ & $-L_3$ & $-L_3$ & $-L_3$ & $-L_1$
& $-L_3$ & $-L_3$ & $-L_3$ & $-L_2$ & $-L_1$ &  $L_1$ & $-L_2$ & $-L_2$ & $-L_1$
& $-L_1$ &  $L_1$ & $-L_1$ &  0 &  0 &  $L_1$ &  0 &  $L_1$ &  $L_2$ &  $L_1$ & 
$L_3$ &  $L_3$ \\ \hline
${\mathbf D_{1}}$ & $-L_3$ & $-L_3$ & $-L_3$ & $-L_3$ & $-L_3$ & $-L_3$ &  0 &
$-L_3$ & $-L_3$ & $-L_3$ & $-L_2$ & $-L_2$ &  $L_1$ & $-L_2$ & $-L_1$ & $-L_1$ &
 0 &  $L_2$ & $-L_1$ &  0 &  0 &  $L_2$ &  0 &  $L_1$ &  $L_2$ &  $L_1$ &  $L_3$
&  $L_3$ \\ \hline
${\mathbf D_{2}}$ & $-L_3$ & $-L_3$ & $-L_3$ & $-L_3$ & $-L_3$ & $-L_3$ & $-L_1$
& $-L_3$ & $-L_3$ & $-L_2$ & $-L_2$ & $-L_2$ &  $L_1$ & $-L_2$ & $-L_1$ & $-L_1$
&  0 &  $L_1$ & $-L_1$ &  0 &  0 &  $L_3$ &  0 &  $L_1$ &  $L_3$ &  $L_1$ & 
$L_3$ &  $L_3$ \\ \hline
${\mathbf D_{3}}$ & $-L_3$ & $-L_3$ & $-L_3$ & $-L_3$ & $-L_3$ & $-L_3$ & $-L_1$
& $-L_3$ & $-L_3$ & $-L_2$ & $-L_2$ & $-L_1$ &  $L_2$ & $-L_2$ & $-L_1$ & $-L_1$
&  0 &  $L_2$ & $-L_1$ &  0 &  0 &  $L_2$ &  0 &  $L_1$ &  $L_3$ &  $L_1$ & 
$L_3$ &  $L_3$ \\ \hline
${\mathbf D_{4}}$ & $-L_3$ & $-L_3$ & $-L_3$ & $-L_3$ & $-L_3$ & $-L_3$ & $-L_1$
& $-L_3$ & $-L_3$ & $-L_3$ & $-L_1$ & $-L_1$ &  $L_1$ & $-L_2$ & $-L_2$ & $-L_1$
& $-L_1$ &  $L_2$ & $-L_1$ &  0 &  0 &  $L_2$ &  0 &  $L_1$ &  $L_2$ &  $L_1$ & 
$L_2$ &  $L_3$ \\ \hline
${\mathbf D_{5}}$ & $-L_3$ & $-L_3$ & $-L_3$ & $-L_3$ & $-L_3$ & $-L_3$ &  0 &
$-L_3$ & $-L_3$ & $-L_3$ & $-L_1$ & $-L_1$ &  $L_1$ & $-L_2$ & $-L_2$ & $-L_1$ &
$-L_1$ &  $L_2$ & $-L_1$ &  0 &  0 &  $L_2$ &  0 &  $L_1$ &  $L_2$ &  $L_1$ & 
$L_2$ &  $L_3$ \\ \hline
${\mathbf D_{6}}$ & $-L_3$ & $-L_3$ & $-L_3$ & $-L_3$ & $-L_3$ & $-L_3$ & $-L_1$
& $-L_3$ & $-L_3$ & $-L_3$ & $-L_2$ & $-L_1$ &  $L_1$ & $-L_2$ & $-L_2$ & $-L_1$
&  $L_1$ &  $L_2$ & $-L_1$ &  0 &  $L_1$ &  $L_2$ &  0 &  $L_1$ &  $L_2$ & 
$L_1$ &  $L_2$ &  $L_3$ \\ \hline
${\mathbf D_{7}}$ & $-L_3$ & $-L_3$ & $-L_3$ & $-L_3$ & $-L_3$ & $-L_3$ & $-L_1$
& $-L_3$ & $-L_3$ & $-L_3$ & $-L_3$ & $-L_1$ &  $L_1$ & $-L_2$ & $-L_2$ & $-L_1$
& $-L_1$ &  $L_1$ & $-L_1$ & $-L_1$ &  0 &  $L_3$ &  0 &  $L_1$ &  $L_3$ & 
$L_2$ &  $L_3$ &  $L_3$ \\ \hline
${\mathbf D_{8}}$ & $-L_3$ & $-L_3$ & $-L_3$ & $-L_3$ & $-L_3$ & $-L_3$ &  0 &
$-L_3$ & $-L_3$ & $-L_3$ & $-L_3$ & $-L_1$ &  $L_1$ & $-L_2$ & $-L_1$ & $-L_1$ &
 0 &  $L_2$ & $-L_1$ &  0 &  0 &  $L_2$ &  $L_1$ &  $L_1$ &  $L_2$ &  $L_3$ & 
$L_3$ &  $L_3$ \\ \hline\hline

${\mathbf D_{10}}$ & $-L_3$ & $-L_3$ & $-L_3$ & $-L_3$ & $-L_2$ & $-L_2$ &  0 &
$-L_3$ & $-L_3$ & $-L_3$ & $-L_2$ & $-L_1$ &  $L_2$ & $-L_3$ & $-L_2$ & $-L_1$ &
$-L_1$ &  $L_2$ & $-L_1$ &  0 &  $L_1$ &  $L_2$ &  $L_1$ &  $L_1$ &  $L_3$ & 
$L_1$ &  $L_3$ &  $L_3$ \\ \hline
${\mathbf D_{11}}$ & $-L_3$ & $-L_3$ & $-L_3$ & $-L_3$ & $-L_3$ & $-L_2$ &
$-L_1$ & $-L_3$ & $-L_3$ & $-L_1$ & $-L_1$ & $-L_1$ &  $L_1$ & $-L_3$ & $-L_1$ &
 0 &  0 &  $L_2$ & $-L_1$ &  $L_1$ &  $L_2$ &  $L_3$ &  $L_2$ &  $L_3$ &  $L_3$
&  $L_3$ &  $L_3$ &  $L_3$ \\ \hline
${\mathbf D_{12}}$ & $-L_3$ & $-L_3$ & $-L_3$ & $-L_3$ & $-L_3$ & $-L_2$ &  0 &
$-L_3$ & $-L_3$ & $-L_3$ & $-L_2$ &  0 &  $L_2$ & $-L_3$ & $-L_3$ &  0 &  $L_1$
&  $L_2$ & $-L_1$ &  $L_1$ &  $L_2$ &  $L_3$ &  $L_1$ &  $L_2$ &  $L_3$ &  $L_2$
&  $L_3$ &  $L_3$ \\ \hline
${\mathbf D_{13}}$ & $-L_3$ & $-L_3$ & $-L_3$ & $-L_3$ & $-L_3$ & $-L_3$ &
$-L_1$ & $-L_3$ & $-L_3$ & $-L_2$ & $-L_2$ &  0 &  $L_1$ & $-L_3$ & $-L_2$ &
$-L_2$ &  0 &  $L_2$ & $-L_2$ &  0 &  $L_2$ &  $L_2$ &  $L_2$ &  $L_2$ &  $L_3$
&  $L_3$ &  $L_3$ &  $L_3$ \\ \hline
${\mathbf D_{14}}$ & $-L_3$ & $-L_3$ & $-L_3$ & $-L_3$ & $-L_3$ & $-L_2$ &
$-L_1$ & $-L_3$ & $-L_3$ & $-L_2$ & $-L_2$ &  0 &  $L_1$ & $-L_2$ & $-L_2$ &
$-L_1$ &  $L_1$ &  $L_2$ & $-L_2$ & $-L_1$ &  $L_1$ &  $L_2$ &  0 &  $L_2$ & 
$L_3$ &  $L_3$ &  $L_3$ &  $L_3$ \\ \hline
${\mathbf D_{15}}$ & $-L_3$ & $-L_3$ & $-L_3$ & $-L_2$ & $-L_2$ & $-L_2$ &
$-L_1$ & $-L_3$ & $-L_3$ & $-L_2$ & $-L_2$ &  0 &  $L_2$ & $-L_3$ & $-L_2$ &
$-L_2$ &  $L_1$ &  $L_2$ & $-L_2$ & $-L_1$ &  $L_1$ &  $L_3$ &  0 &  $L_2$ & 
$L_3$ &  $L_3$ &  $L_3$ &  $L_3$ \\ \hline
${\mathbf D_{16}}$ & $-L_3$ & $-L_3$ & $-L_3$ & $-L_3$ & $-L_3$ & $-L_3$ &
$-L_1$ & $-L_3$ & $-L_3$ & $-L_3$ & $-L_3$ & $-L_2$ &  $L_1$ & $-L_3$ & $-L_3$ &
 0 &  $L_1$ &  $L_1$ & $-L_1$ &  $L_1$ &  $L_1$ &  $L_2$ &  $L_1$ &  $L_1$ & 
$L_2$ &  $L_2$ &  $L_2$ &  $L_3$ \\ \hline
${\mathbf D_{17}}$ & $-L_3$ & $-L_3$ & $-L_3$ & $-L_2$ & $-L_2$ & $-L_1$ &  0 &
$-L_3$ & $-L_3$ & $-L_2$ & $-L_2$ & $-L_1$ &  $L_2$ & $-L_3$ & $-L_2$ & $-L_1$ &
 $L_1$ &  $L_2$ & $-L_2$ &  $L_1$ &  $L_1$ &  $L_3$ &  $L_1$ &  $L_2$ &  $L_3$ &
 $L_2$ &  $L_3$ &  $L_3$ \\ \hline
\end{tabular}}
\end{center}
\end{table}

\begin{figure}[p]
\centering
\includegraphics[width=0.6\columnwidth]{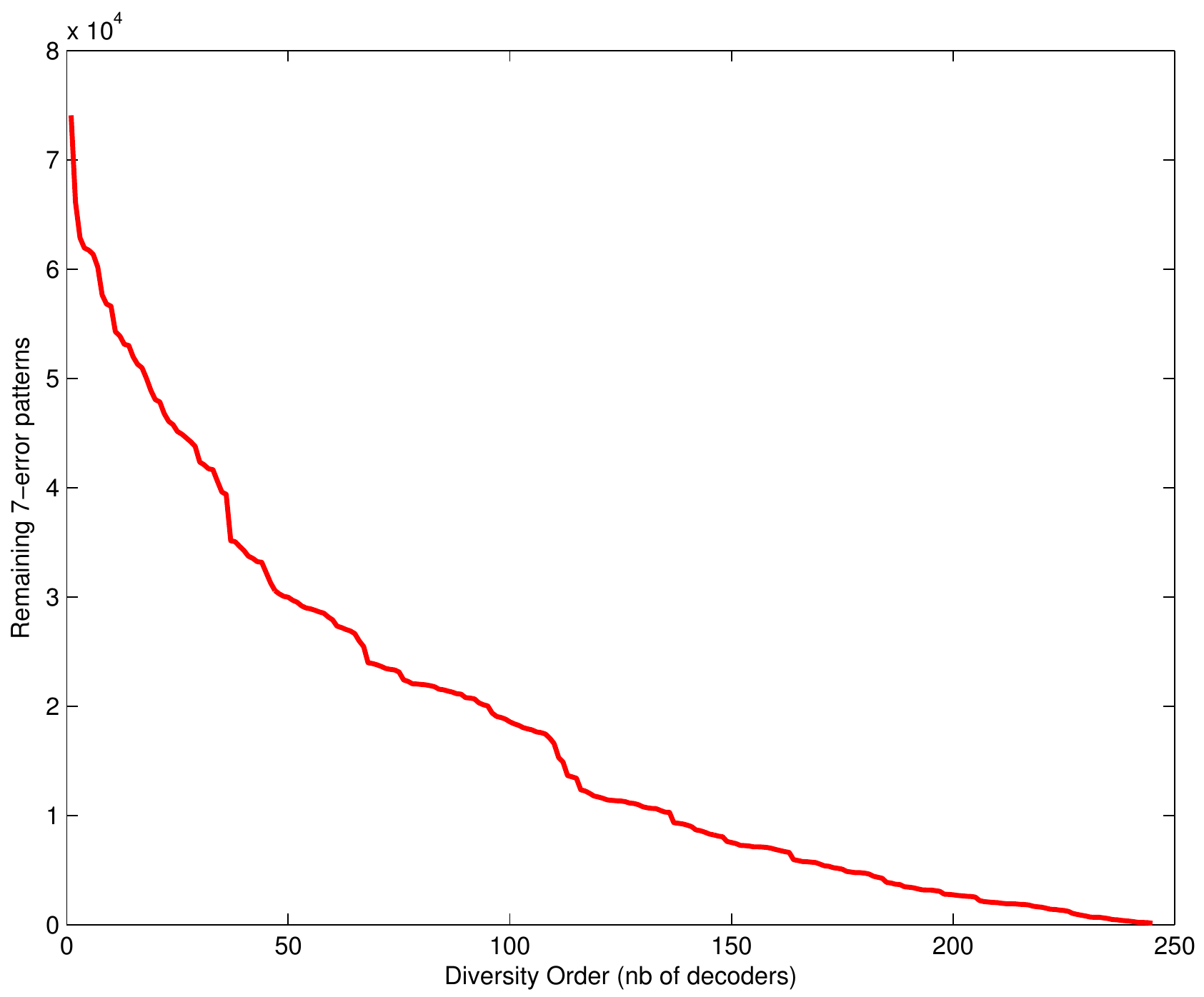}
\caption{Number of remaining uncorrected $7$-error patterns with sequential use
of FAIDs in the diversity sets.}
\label{fig:Plot_7errors}
\end{figure}

\end{document}